# The Jive Verification System and its Transformative Impact on Weather Forecasting Operations

Nicholas Loveday,[a] Deryn Griffiths,[a] Tennessee Leeuwenburg,[a] Robert Taggart,[b] Thomas C. Pagano,[a] George Cheng,[a] Kevin Plastow,[c] Elizabeth Ebert,[a] Cassandra Templeton,[a] Maree Carroll,[a] Mohammadreza Khanarmuei,[c] Isha Nagpal [a]

[a] *Bureau of Meteorology, Melbourne, Victoria, Australia*

[b] *Bureau of Meteorology, Sydney, New South Wales, Australia*

[c] *Bureau of Meteorology, Brisbane, Queensland, Australia*

nicholas.loveday@bom.gov.au

ABSTRACT

Forecast verification is critical for continuous improvement in meteorological organizations. The Jive verification system was originally developed to assess the accuracy of public weather forecasts issued by the Australian Bureau of Meteorology. It started as a research project in 2015 and gradually evolved to be a Bureau operational verification system in 2022. The system includes daily verification dashboards for forecasters to visualize recent forecast performance and "Evidence Targeted Automation" dashboards for exploring the performance of competing forecast systems. Additionally, Jive includes a Jupyter Notebook server with the Jive Python library which supports research experiments, case studies, and the development of new verification metrics and tools.

This paper describes the Jive verification system and how it helped bring verification to the forefront at the Bureau of Meteorology, leading to more accurate, streamlined forecasts. Jive has provided evidence to support forecast automation decisions and has helped to understand the evolving role of meteorologists in the forecast process. It has given operational meteorologists tools for evaluating forecast processes, including identifying when and how manual interventions lead to superior predictions. Work on Jive led to new verification science, including novel metrics that are decision-focused, including diagnostics for extreme conditions. Jive also provided the Bureau with an enterprise-wide data analysis environment and has prompted a clarification of forecast definitions.

These collective impacts have resulted in more accurate forecasts, ultimately benefiting society, and building trust with forecast users. These positive outcomes highlight the importance of meteorological organizations investing in verification science and technology.

## Introduction

In recent decades, verification research and development have been a focal point for the scientific community (Casati et al. 2008; Ebert et al. 2013). Verification of forecasts against meteorological observations provides the quantitative basis for understanding the accuracy, characteristics, and value of forecasts, and for assessing whether one system is better than another. It is also essential for demonstrating how forecast systems can be improved and what role operational meteorologists play in the forecast process, given the continuous improvement in modelling capability.

Suitable software packages are critical to support scientists so that they can efficiently verify large numbers of forecasts. Several verification packages have emerged over the last two decades (e.g., Brown et al. 2010; NCAR - Research Applications Laboratory 2014; Eyring et al. 2016; Manubens et al. 2018; Righi et al. 2020; Siegert et al. 2020; Lauer et al. 2020; Weigel et al. 2021; Brady and Spring 2021; Brown et al. 2021; Jordan et al. 2019; Gilleland 2022; Zanetta and Allen 2024) to support operational meteorological organizations and research institutions.

In 2015, due to the lack of appropriate tools at the time to assess public weather forecasts, the Australian Bureau of Meteorology (hereafter "the Bureau") initiated the development of a new verification system called Jive. The primary objective was to evaluate the Bureau's public weather forecasts (as distinct from its warnings, observations, or specialized services). This paper describes the Python-based verification system, Jive, and its utilization in the Bureau. We discuss the background and motivation for developing Jive and its positive impact in supporting decisions that have led to more accurate and streamlined forecasts in the Bureau. The examples that we present emphasize the need for meteorological agencies to strengthen both their verification science and their technology resources, as well as integrate verification activities into the daily workflows of operational meteorologists. Finally, we discuss how verification supports the adoption of more automation and can be used to answer questions about the role of the operational meteorologist in the forecast production process.

## History and overview of the Jive verification system

### *The forecast production process and the need for Jive*

The need to verify the Bureau's public weather forecasts to understand their quality prompted the inception of Jive to replace an older verification codebase that was difficult to use and support. An overview of the forecast process at the Bureau is shown in Fig. 1. Initially Jive was created to provide easy access to forecast and observational data with a range of statistical analysis tools to conduct verification on forecasts produced using the Graphical Forecast Editor (GFE). Developed originally by the U.S. National Weather Service and later adopted by the Bureau, the GFE is a forecasting platform that visualizes Numerical Weather Prediction (NWP) and model-based consensus guidance. The GFE also provides graphical editing tools that allow operational meteorologists to manually adjust automated gridded forecasts. It includes automated processes that produce a set of grids known as AutoFcst (Griffiths and Jayawardena 2022), that are an automated alternative to the official human-curated gridded forecasts. Additionally, it contains natural language generation to produce automated text forecasts, and the ability to disseminate forecasts (Leeuwenburg 2009). The forecasts and warnings produced through the GFE (hereafter referred to as “official” forecasts) are disseminated to diverse users via many channels including the Bureau's website, mobile app, and the Australian Digital Forecast Database (ADFD) (BOM 2015). The initial implementation of Jive was primarly to help the Bureau understand how the GFE forecast process could be streamlined to improve predictive performance and production efficiency, with a particular focus on differences between automated forecasts and those manually curated by operational meteorologists (Just and Foley 2020).

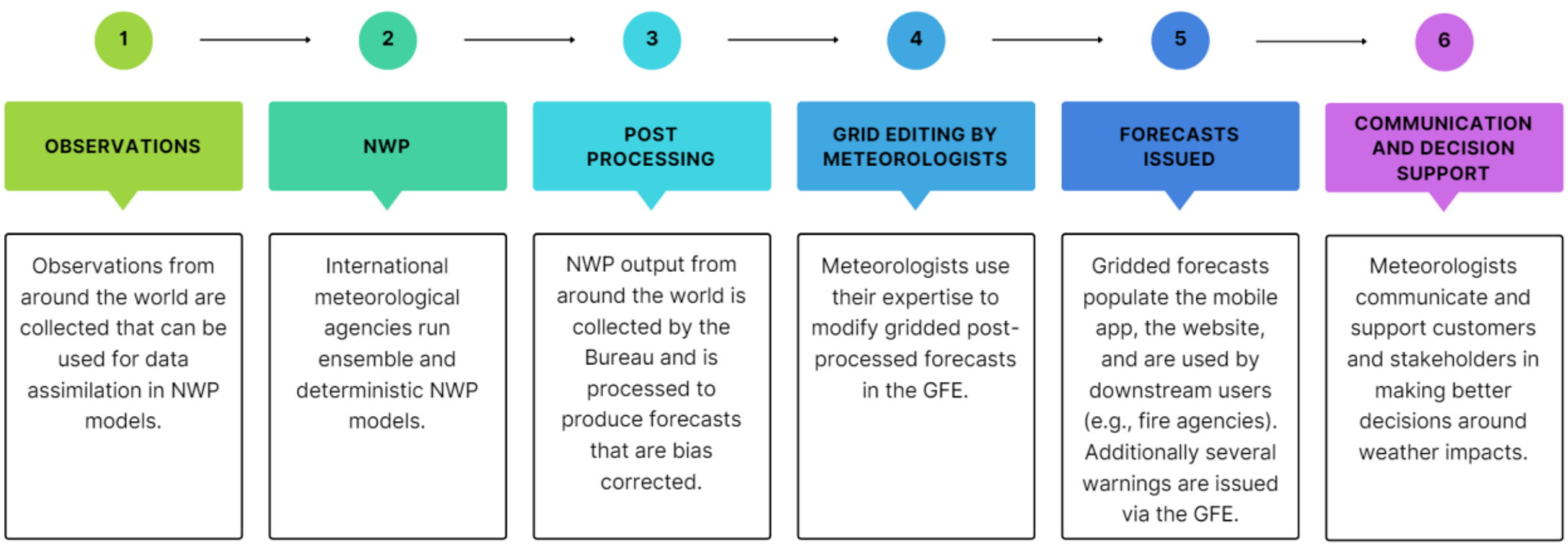

Figure 1. An overview of the steps to produce the official public weather forecasts at the Bureau. Jive was originally designed to evaluate and support steps 3-5.

At Jive's inception, most existing verification software packages focused primarily on NWP verification (step 2 in Fig 1) and were less suited to verification of official Bureau forecasts (step 5 in Fig 1). Being able to flexibly verify N-dimensional data is important for evaluating different types of forecasts using the one tool. For example, a forecast timeseries at a single point is a 1D array, while a gridded forecast, with 3 spatial dimensions (*x*, *y*, *z*), with multiple lead times, and multiple parameters would be a 5D array. There were no publicly available Python verification packages that supported in-memory calculations on N-dimensional labeled data, that allowed new user-focused metrics to be easily added by scientists.

### *An operational verification system with interactive tools*

While beginning as a small research system, Jive evolved to become a supported operational system in 2022 for verifying public weather forecasts. Its key components are summarized in Fig. 2. It features daily updated "Forecaster Verification Dashboards" for operational meteorologists to visualize recent forecast performance and "Evidence Targeted Automation" (ETA) dashboards for exploring the performance of competing forecast systems across several months. Additionally, Jive provides a Jupyter Notebook server with the Jive Python library to support experimentation, tailored case studies, and the development of novel verification metrics and tools in a web browser. Several graphics from these dashboards and Jupyter Notebooks will be shown in the following sections. Providing all Bureau staff with access to verification dashboards and Jupyter Notebooks with a Python verification library has been critical in supporting effective Research to Operations and Operations to Research exchanges within the Bureau. Data, Jupyter Notebooks, and metrics in Jive are increasingly being used to support research (Griffiths et al. 2017, 2019, 2021; Foley and Loveday 2020; Just and Foley 2020; Price and Foley 2021; Hurn et al. 2021; Taggart 2022a,b,c; Taggart et al. 2022; Taggart 2023; Loveday et al. 2024).

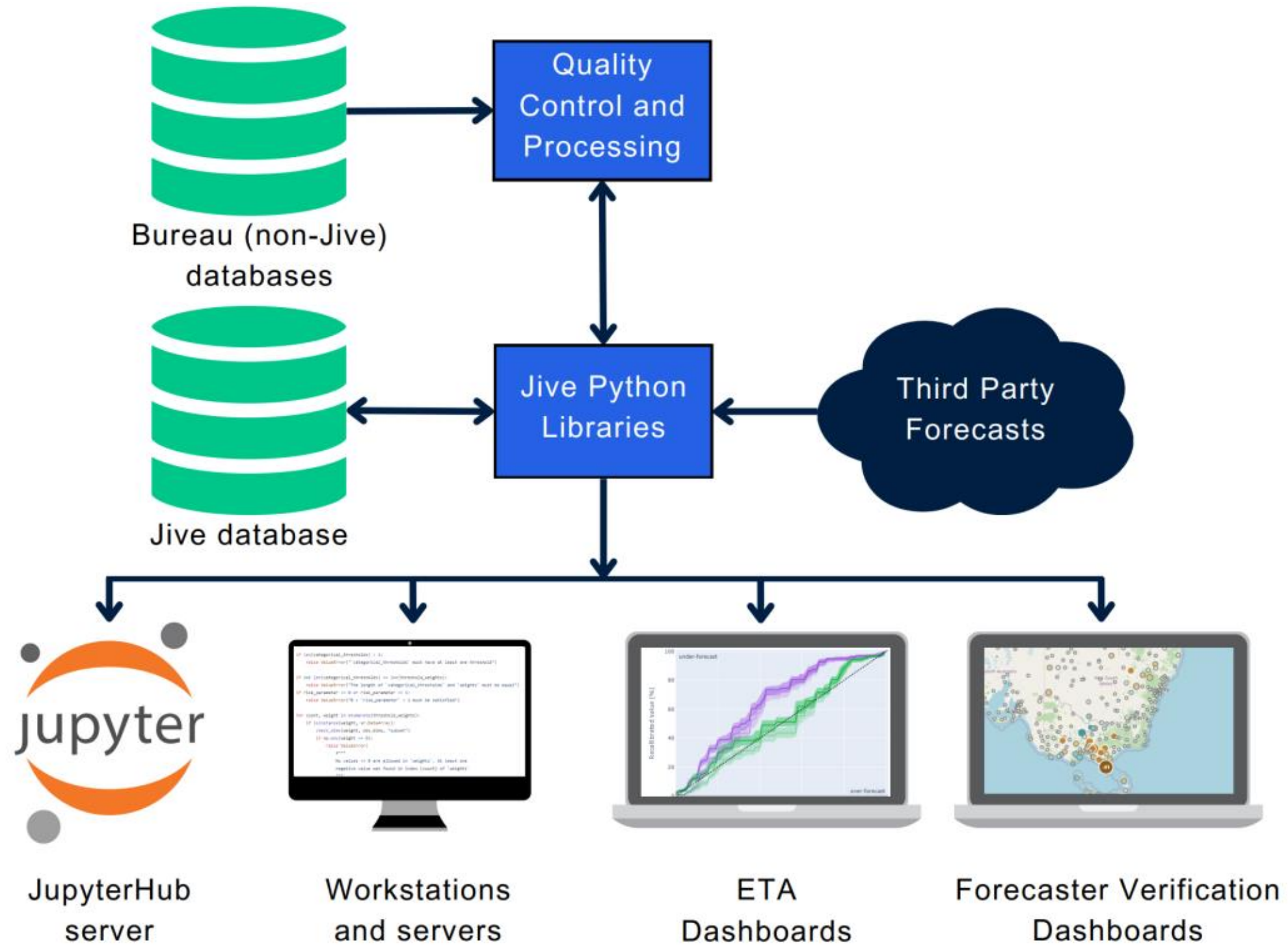

Figure 2. Schematic diagram of the Jive system. Bureau databases supply forecasts and raw observations. Integrated Jive Python libraries do everything from data retrieval, processing, and storage, as well as generate statistics and visualize outputs on dashboards.

### *Jive database*

The Jive database stores both gridded and point-based forecast and observation data. Forecast datasets include the official forecasts as well as several automated alternatives. Forecasts for the Australian region from third parties are also ingested and made available for Jive users. Table A1 in Appendix A provides a summary of the forecast parameters stored in the Jive database. Point-based observation data is sourced from one-minute Automatic Weather Stations (AWS) and wave buoys. These observations undergo quality control and are processed so that they match the forecast service definitions (e.g., the maximum 10-minute mean wind during an hour). Lightning sensor data and analysis grids are also used as observations. Table A2 in Appendix A provides an overview of observations in the Jive database.

During the development of Jive, operational meteorologists informed us that they would trust verification results more if AWS observations were used instead of gridded analyses. This is because the available gridded analyses had several known issues with their quality, and typically did not align with forecast service definitions. For example, gridded rainfall analyses were aggregated across 24-hour periods to 9 am local time at each location, while rainfall forecasts were for 24-hour and 3-hour periods aligned with 1500 UTC (12:30 am ACST). Moreover, the forecasts are commonly used by users at a point in space and AWS provide data on a more relevant spatial scale than the area-averaged grid cell from a gridded analysis. For these reasons, the database has a large collection of forecasts at grid points nearest to AWS locations and a smaller collection of complete gridded forecasts. Groups of selected weather stations were defined for the verification for each meteorological parameter. To reduce the impact of representativeness issues, stations not representative of the forecast grid cell were excluded from each group (e.g., stations strongly affected by sub-grid local effects such as weather stations placed on the edge of an exposed cliff).

### *Metrics and statistical tools*

Jive contains a rich variety of metrics including widely used verification metrics, novel metrics developed by Bureau scientists, and other diagnostics such as Murphy Diagrams[1] (Ehm et al. 2016) which, to our knowledge, lack a publicly available Python implementation. These metrics have been developed in Jive to flexibly handle forecasts and observations with varying sets of dimensions, enabling vectorized calculations to work efficiently on both point-based and gridded data across both multiple lead times and multiple valid times.

The selection of metrics was chosen primarily for their utility to verify public weather forecasts based on statistically defined, post-processed NWP forecasts rather than verifying NWP models using spatial methods (e.g., Gilleland et al. 2009). Point-based verification metrics were used because the Bureau aims to optimize these forecasts at the user scale, i.e., for users at point locations. The metrics in Jive can evaluate single-valued continuous forecasts (e.g., expected values, or quantile forecasts), categorical forecasts, and interval forecasts (e.g., "5mm to 10mm of rainfall", where the values are the 25th and 75th percentile forecasts respectively). Moreover, Jive includes metrics for evaluating probabilistic forecasts which can be expressed as a cumulative distribution function, an ensemble, or the probability of a binary outcome.

There has been a strong focus on implementing user-focused metrics and tools. Examples include

a) metrics that measure value such as Murphy Diagrams (Ehm et al. 2016) and Relative Economic Value (Richardson 2000),

---

[1] Murphy Diagrams show the mean score as a function of decision threshold. They show the benefit provided by using the forecast to predict whether the outcome will be above or below a decision threshold. We have found them particularly useful for visualising forecast performance for extreme thresholds and comparing how the performance of different forecast systems differ from each other at important user-decision thresholds.

b) threshold weighted scores for forecasts expressed as predictive distributions (Gneiting and Ranjan 2011), as a single value (Taggart 2022a), or as categories (Taggart et al. 2022), and

c) diagnostic tools such as isotonic regression (Dimitriadis et al. 2021).

Jive can also calculate confidence intervals using methods such as the one-tailed Student's t-test, binomial confidence intervals, and the modified Diebold-Mariano Test (Diebold and Mariano 1995; Hering and Genton 2011).

### *A flexible verification tool*

Xarray is a python library that handles multidimensional, labelled arrays of data (Hoyer and Hamman 2017). Jive uses xarray to manage its data which allows for a flexible, interactive experience for Jive users. Since metrics in Jive are implemented to handle multiple dimensions, it is possible to use the same metric on gridded, point-based, or area-aggregated data. Handling data in-memory avoids bottlenecks in reading and writing data and allows the verification tool to be highly interactive. For example, a user can load data into a Jupyter Notebook, perform pre-processing, compute verification metrics, and generate plots all through a web browser without the need to write data to disk. For computations on large datasets surpassing memory constraints, Dask (Rocklin 2015) can be used, enabling Jive to scale with parallel computations. Additionally, Jive leverages xarray's ability to analyze data stored in most map projections. Jive can be used in various ways to support various users with a range of technical and scientific needs and expertise. Table 1 summarizes the purpose and target audience of the four primary ways of using Jive.

| **Mode of use** | **Target Audience** | **Purpose** |
|---|---|---|
| Jupyter Notebooks via the JupyterHub server | Scientists who want an environment where they can easily write Python code with access to Jive data, Jive metrics, and scientific Python libraries. Additionally, they may be shared with staff who do not write Python code. | Case studies, developing grid-editing forecast strategies, developing new verification methods, building prototype verification solutions that could be converted into dashboards, and a general data analytics environment for the organization. |
| Workstations and servers | Scientists and software developers who want to use aspects of Jive, but need more control or features than are available in Jupyter Notebooks | The ability to retrieve forecast and observation data from the Jive database for use outside of the Jive virtual machines. Development of standalone systems that use Jive, such as systems that generate forecast performance KPIs. |
| Evidence Targeted Automation Dashboards | Operational meteorologists, verification scientists, and managers. No ability to write code is required to use these dashboards. | Forecast system comparison with a focus on providing data to support informed decisions about automation. |
| Forecaster Verification Dashboards | Operational meteorologists and verification scientists | Daily review of forecast performance, developing grid editing strategies, and post-event review. |

Table 1. A description of the target audience and purpose of the four primary ways to use Jive as shown on the lower row of Fig. 2.

Additionally, the metrics from Jive are being migrated to the open-source Python library *scores*, which will be discussed later in this paper. This will allow anyone to be able to use the metrics component of the Jive system in isolation from the rest of the system.

## Supporting decisions to automate the production of weather forecasts

### *Evidence Targeted Automation dashboards*

An aim in developing Jive was to help inform the best approach for streamlining the forecasting process, optimizing its mix of automated forecasts and manual revision. To achieve this, a set of interactive dashboards, known as the ETA dashboards, were created that compare the accuracy, economic value, and characteristics of fully automated forecasts with official forecasts prepared by operational meteorologists. These dashboards expand on the framework for supporting decisions around the automation of forecasts set out in Griffiths et al. (2017).

The ETA dashboards display plots of spatially and temporally aggregated verification scores and diagnostics for competing forecasts across different lead times. When the user selects a region of interest from the menu, an interactive map shows which AWSs were included in the verification. Additionally, a dynamically updating “plot explanation” tool describes how to interpret the verification results, aiding managers or meteorologists who may not be verification experts in correctly interpreting the plots on the dashboard.

These ETA dashboards have provided the Bureau with an understanding of where operational meteorologists both do and do not add value or improve accuracy in the production of forecasts in the GFE. They have also shown the accuracy of the automated forecasts (AutoFcst) and have motivated and informed improvements to automated guidance.  As verification instilled greater confidence in the overall quality of AutoFcst among operational meteorologists and managers, the forecast process has become more streamlined over time (Just and Foley 2020). This increase in automation has led to forecasts that are spatially consistent across the country, with more consistent forecast characteristics across lead times, and improved accuracy in many situations. It has also freed up time for operational meteorologists to participate more actively in decision-support activities.

### *Improving and clarifying the forecast definition*

Adoption of automated forecasts by operational meteorologists is inhibited when automated forecasts have characteristics that do not reflect forecast objectives valued by meteorologists. Verification highlighted these differences in forecast objectives, some of which were resolved by more clearly defining some of the Bureau’s gridded forecasts.

For example, many possible kinds of forecasts could be made for a parameter such as hourly wind speed. If someone requests an hourly wind speed forecast as a single number, they may want a middle value (like the median or mean) or an upper value (like the 90th percentile) of all possible outcomes. They may want to know about wind speeds that are predicted to occur on the hour, the maximum within the hour, or something else. Without a clear forecast directive or target, the forecaster does not know what value to predict, and it is also unclear which metric should be chosen to assess forecast performance.

The Bureau's wind speed forecasts were ill-defined for many years. The automated forecasts were calibrated to forecast a median-value sustained wind for each timestamp. Meanwhile, operational meteorologists produced forecasts that aligned closer with the peak sustained wind within each 60-minute period, as the latter was more suitable for deriving wind warnings and some fire danger products.

The ETA wind speed dashboards clearly highlighted these differences, prompting a discussion to clarify the forecast definition. Consequently, it was decided to define the wind speed forecast as the expected 10-minute mean wind speed at the timestamp ("wind-on-hour") and introduce a new gridded forecast that targeted the expected peak 10-minute mean wind within 30 minutes either side of the timestamp ("max-wind-in-hour"). A strong linear relationship between wind-on-hour and max-wind-in-hour speeds was found using least squares regression on observations. This relationship was used to predict max-wind-in-hour speeds from forecast wind-on-hour speeds in a hindcast experiment using Jive.  The hindcast forecasts significantly outperformed the official and old automated forecasts in its prediction of maximum 10-minute mean winds within each hour across all lead days. The MSE of these three forecast systems is shown in Fig 3. While many verification methods were used to investigate the difference in quality between different forecast sources, we mainly summarize the performance of the forecast sources in this article using the MSE which is consistent with forecasting the mean (i.e., expected) value (Gneiting 2011).

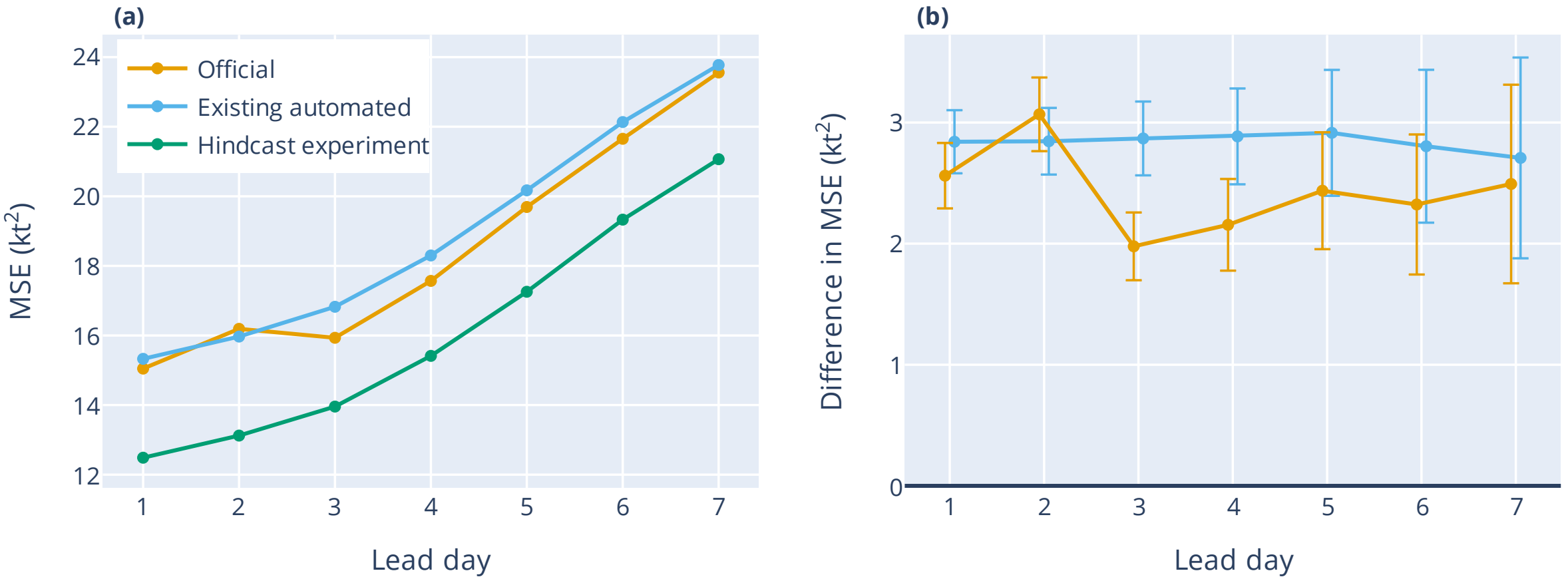

Figure 3. **a)** The Mean Squared Error (MSE) of the official forecasts (orange), the automated forecasts available to meteorologists at the time (blue), and the hindcast experiment (green) are displayed for each lead day[2]. Wind speed forecasts are evaluated against the maximum wind-in-hour observations for all standard height AWSs across Australia (see Appendix B for a map of AWS locations) for Spring (September-November) 2019. **b)** The difference in MSE is shown for official minus the hindcast (orange) and automated minus the hindcast (blue) with 95% confidence intervals[3]. Positive values indicate the hindcast performed better than the forecast it was compared against. Note that the automated forecast guidance available at the time only predicted wind speed on-the-hour rather than the maximum in the hour.

This linear relationship has since been implemented into the GFE tools for use by operational meteorologists and AutoFcst to produce the two gridded wind speed forecasts. Since implementing these changes, verification has demonstrated an improvement in accuracy and consistency of forecast characteristics across all lead times.

*Improving automated guidance by understanding forecast strategies used by operational meteorologists*

---

[2] In each example in this paper, forecast and observations have been matched in the sense that if data is missing for a location and timestep for one forecast system, that point will also be treated as missing for the other forecast system when verification results are calculated.

[3] Error bars around the difference in mean error in this paper are based on the Diebold Mariano test statistic (Diebold and Mariano 1995) with the Hering and Genton (2011) modification after accounting for spatial dependency.

Verification highlighted where meteorologists were systematically producing more accurate forecasts than automated guidance. For instance, the ETA dashboards revealed that operational meteorologists were making significant improvements to the automated guidance for wind speed forecasts at high mountain peaks. We interviewed operational meteorologists to understand the grid editing techniques they employed, then applied similar techniques in an automated way via a hindcast experiment. Verification results from that experiment (summarized by the MSE in Fig. 4) supported an upgrade to AutoFcst using those techniques, enabling forecasters to allocate time to other work. The hindcast experiment also showed reduced errors compared to the official forecasts for longer lead times (lead days 5-7), highlighting the benefits of applying the improved automation rather than relying on forecasters to make changes systematically across all lead times.

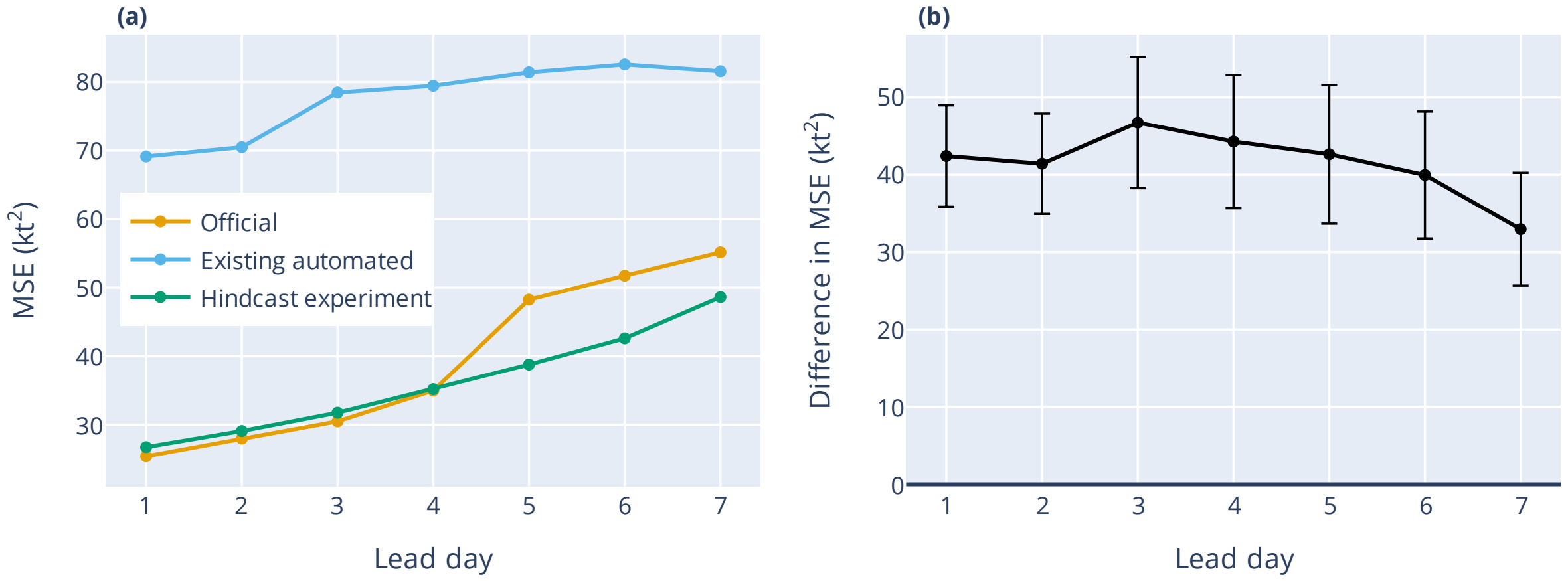

Figure 4. **a)** The MSE of the official forecasts (orange), the automated forecasts available to meteorologists at the time (blue), and the hindcast experiment (green) are displayed for each lead day. Wind speed forecasts are evaluated against wind-on-hour observations from AWS located on high mountain peaks (see Appendix B for a map of AWS locations) during Summer (December-February) 2017-2018. **b)** Differences in the MSE between the existing automated forecasts and the hindcast experiment (automated minus hindcast) with 95% confidence intervals. Positive values indicate the hindcast experiment performed better than the existing automated forecasts.

## Verification tools to support operational meteorologists

### *ETA dashboards used by operational meteorologists to improve on automated forecasts*

The ETA dashboards highlight biases of both automated and official forecasts. Recalibrating (or bias correcting) automated forecasts based on verification is an evidence-based improvement meteorologists can make to the forecasts until such time as the automated guidance is improved.

An instance of this capability is the systematic under-forecast bias in the AutoFcst wind speed forecasts over southwestern parts of Western Australia which was detected on the ETA dashboards. Investigation showed that this bias was caused by non-standard anemometer heights from third-party weather stations contaminating the analysis grid used for calibrating automated forecasts. In response, meteorologists systematically applied bias corrections, guided by verification results displayed on the ETA dashboards, resulting in an improvement of roughly 1-2 lead days of skill for shorter lead times in official forecasts (Fig 5), which is notable, since NWP performance has historically improved by about one lead day of skill per decade (Bauer et al. 2015). This result highlights the ability of humans to improve on automated forecasts if they are aware of deficiencies in the data used to calibrate those forecasts.

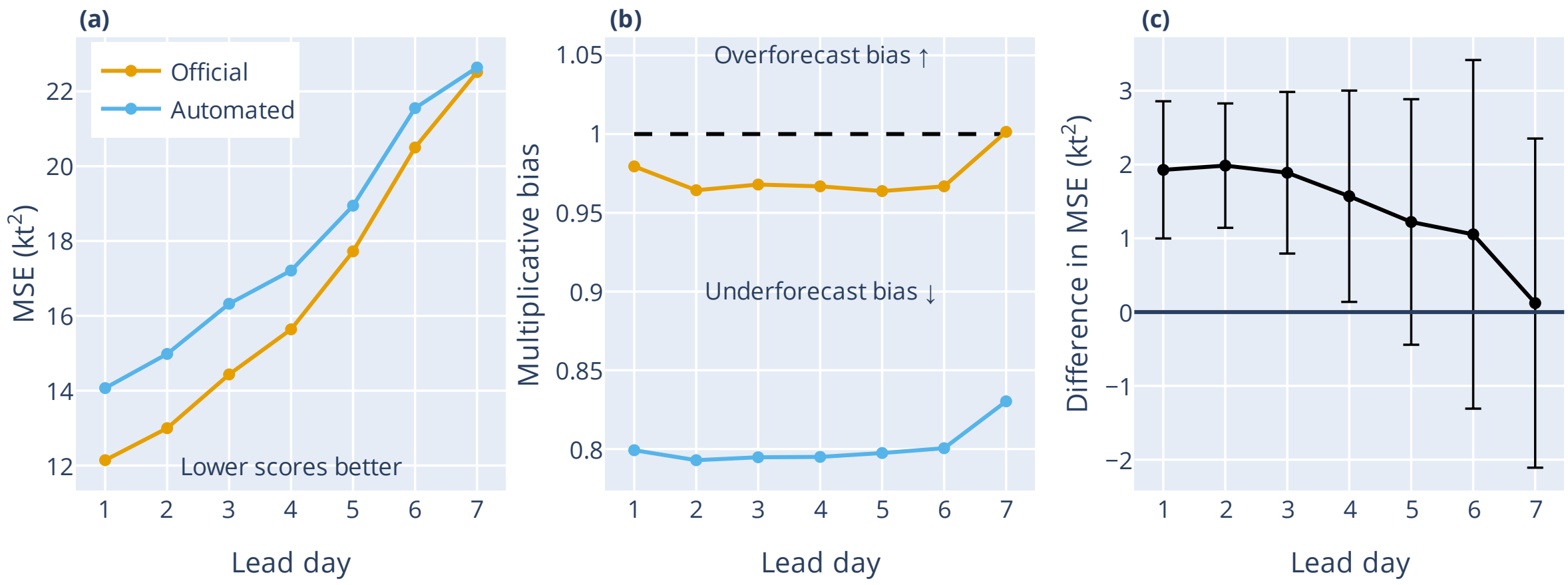

Figure 5. Wind speed performance for the official forecasts (orange) and the automated forecasts (blue) over southwestern Western Australia (see Appendix B for a map of AWS locations) for winter (June-August) 2023 for seven lead days. **a)** MSE. **b)** Multiplicative bias. The dashed black line shows an ideal multiplicative bias of 1. Values below 1 indicate an underforecast bias. **c)** Difference in MSE between official and automated forecasts (automated minus official) with 95% confidence intervals. Positive values indicate that the official forecasts performed better than the automated forecasts.

The ETA dashboards also highlight where operational meteorologists should leave automated forecasts unaltered. For example, during the 2016 dry season over the Northern Territory (May-September) meteorologists were unaware that they were degrading forecast accuracy by modifying the automated forecasts as they did not have access to verification results at the time. Subsequent verification results persuaded operational meteorologists and managers to primarily rely on automated temperature guidance with minimal modifications, thereby augmenting forecast skill by more than a day across shorter lead times in following seasons.

As meteorologists across the country began issuing temperature forecasts to the public primarily based on automated guidance, they were still inclined to adjust the guidance in extreme situations. Most bulk verification statistics masked the differences between the accuracy of the official forecasts and the automated forecasts (Fig. 6a). However, operational meteorologists believed that they were improving on automated guidance in extreme heat situations. To test these beliefs, methods that assess the accuracy of forecasts with an emphasis on extremes while avoiding selection bias caused by conditioning on observed or forecast events were developed (Taggart 2022a), and results were displayed on the ETA dashboards. These results showed that, for example, meteorologists substantially improved on automated forecasts when predicting extreme maximum temperatures over Australia's 2019-2020 "Black Summer" (Mills et al. 2022) (Fig 6b). Results were statistically significant at the 5% level for lead days 1-3. The threshold weighted MSE showed that around 1 lead day of skill was added at shorter lead days, while the traditional MSE did not (compare Fig 6a with Fig 6b).

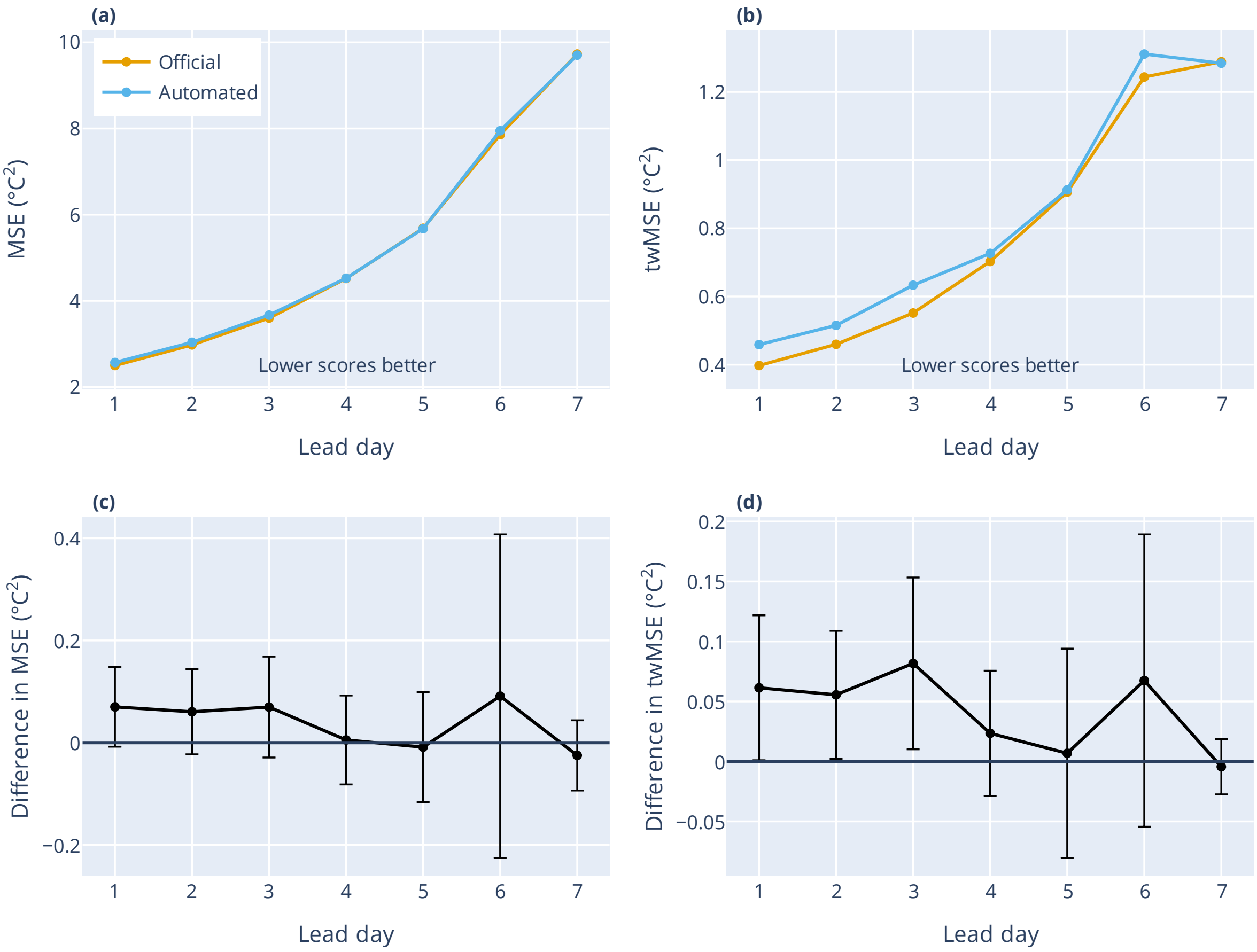

Figure 6. Mean scores of maximum temperature forecast performance for official (orange) and automated (blue) forecasts measured during the 2019-2020 Black Summer (December- February) for all land-based AWS across Australia (see Appendix B for a map of AWS locations). **a)** MSE **b)** Threshold weighted MSE (twMSE) measuring the predictive ability of forecast systems for thresholds above the 97$^{th}$ percentile climatological value at each weather station. **c)** Difference in MSE (Automated minus official) with 95% confidence intervals. Positive values indicate that official forecasts performed better than the automated forecasts. **d)** The same as c), but for twMSE.

### *Forecaster Verification Dashboards*

Operational meteorologists from various forecasting teams in the Bureau have used Jive to develop their own highly interactive, web browser-based, prototype dashboards. One of these prototype dashboards drew data from the Jive database and displayed a core set of essential information intuitively. Scientists and software developers expanded this prototype into a series of operational dashboards known as the "Forecaster Verification Dashboards" (Fig 7.). Users select dates, a weather element, lead day, and a region of interest, after which observations, forecasts, and verification scores are displayed using maps and summary bar charts. Additionally, the dashboards display time series and scatterplots of forecasts and observations as well as mean sea level pressure maps, radar scans, and forecast policy maps to provide context for individual dates.

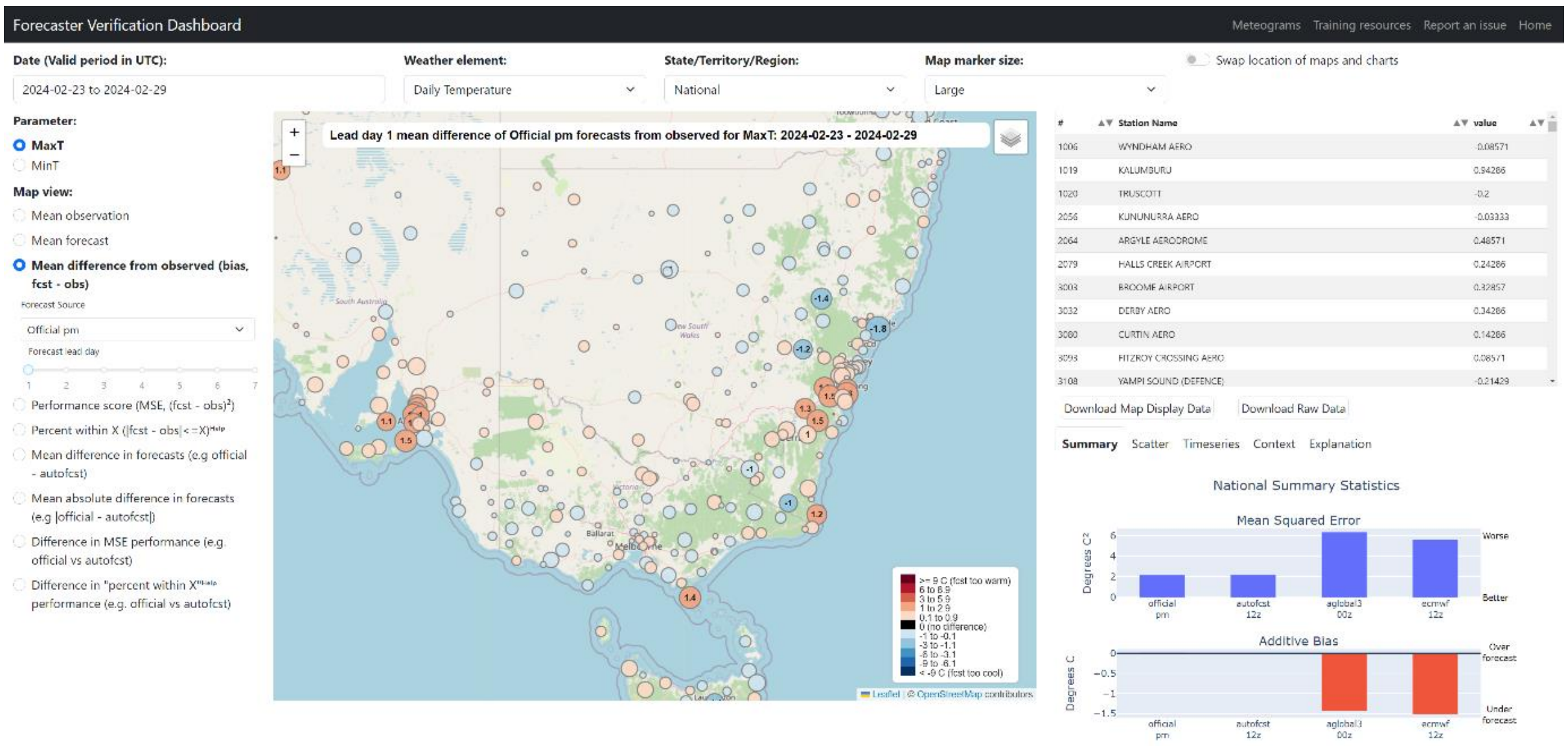

Figure 7. An example of a Forecaster Verification Dashboard. The map shows the biases of the maximum temperature forecasts for the lead day 1 official forecasts for 23 February 2024 to 29 February 2024 at weather station locations across southeastern Australia. Red circles indicate where there was a warm bias, while blue circles show where there was a cold bias. The table to the right shows the bias values at weather stations and the data can be downloaded for further analysis. The lower right bar charts show the MSE and additive bias across Australia for four different forecast systems.

The main audience for these dashboards is operational meteorologists. These dashboards have been embedded into the daily forecast process for operational meteorologists to review how their forecasts, automated forecasts, and NWP models performed over recent days. By examining how forecasts performed at individual sites and dates, operational meteorologists can identify different biases and errors that may occur with certain synoptic conditions. Then they will often recalculate the verification results aggregated over many dates to determine if there is evidence of systematic errors, which can be corrected with improved forecast strategies.

These dashboards have also been used for case studies, post-event reviews, answering media inquiries, and responding to user requests on forecast accuracy.

### *Jive Jupyter Notebooks*

The Jive system includes a JupyterHub server, providing all Bureau staff with the capability to extract forecasts and observations from the Jive database into a Jupyter Notebook equipped with verification metrics and an extensive range of Python libraries. A Jupyter notebook is a tool that allows people to write code, analyze data, and produce visualizations in a web browser. This capability has empowered research scientists to test out innovative ideas and develop new verification metrics. As of February 2024, over 500 Bureau staff have used the JupyterHub server, which is around a quarter of the Bureau's workforce. Notebooks have been used by research scientists and operational meteorologists to conduct verification case studies, perform targeted verification, develop optimal forecast strategies, and design new forecast processes in the GFE.

For example, ETA dashboards highlighted that AutoFcst had an under-forecast bias with its daily probability of precipitation (DailyPoP) forecasts. The GFE has a tool that can recalibrate forecasts using piecewise linear mappings. Operational meteorologists leveraged the Jive Jupyter Notebooks to determine the optimal parameter settings for this tool over distinct geographical areas, and they then systematically applied the tool to recalibrate forecasts, yielding 0.5-1 extra lead day of skill based on the Brier Score (Brier 1950). Figure 8 shows an example of the improvement in the reliability of DailyPoP forecasts that operational meteorologists make when applying this calibration.

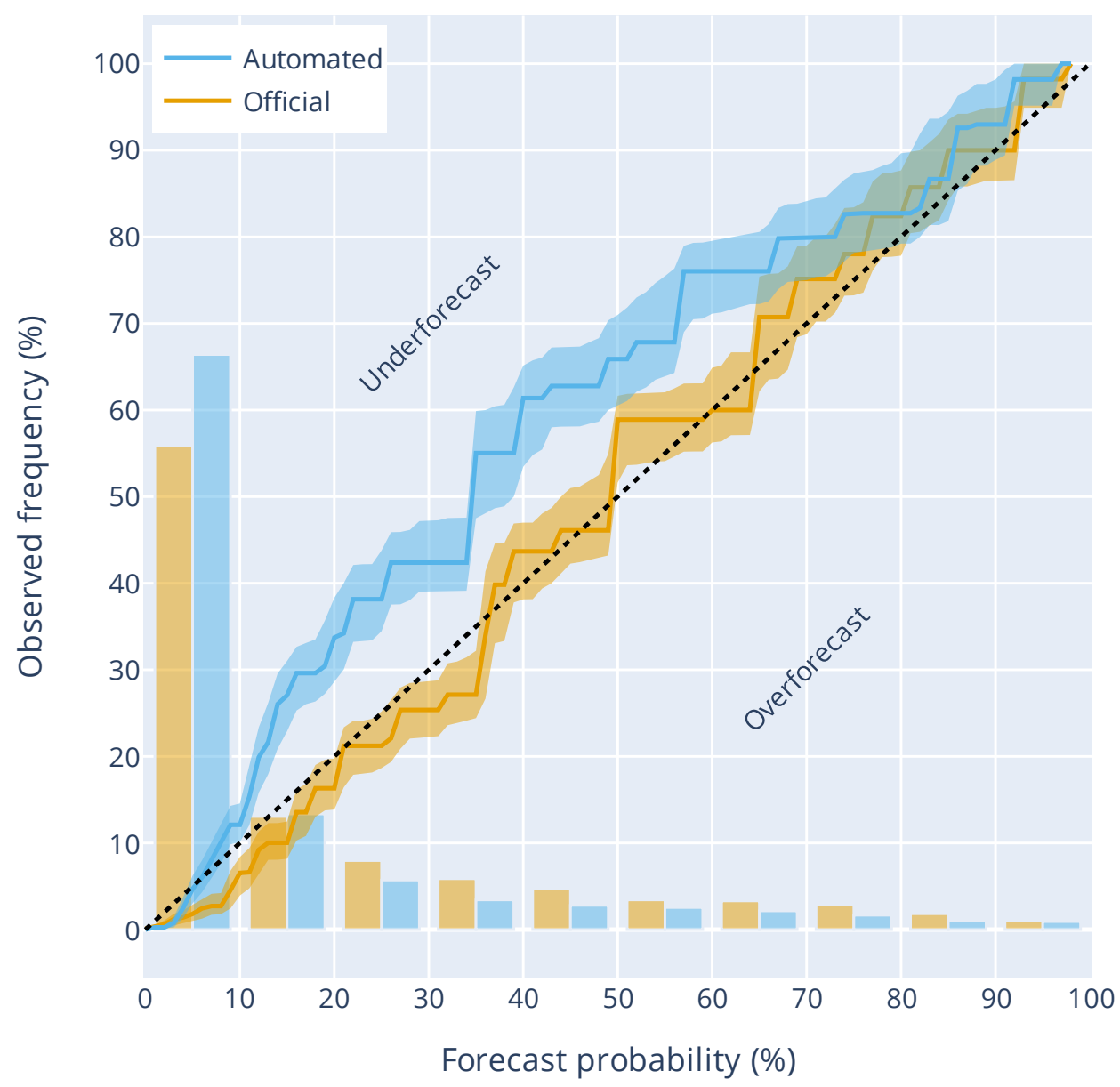

Figure 8. Reliability diagram, based on isotonic regression (Dimitriadis et al. 2021) of DailyPoP forecasts from automated (blue) and the official (orange) forecast systems over inland Australia (see Appendix B for map of AWS) for lead day 1 forecasts for Dec 2021-Feb 2022. 90% confidence bands are generated through 10,000 bootstrap samples. A histogram indicates the frequency of forecasts for probabilities within ten percentage point bins. Forecasts and observations were masked where exactly 0.2mm of rainfall was recorded to remove the impact of spurious recordings due to dew.

## A verification system to complement other tools and support the Bureau

### *Jive and METplus*

The Bureau plans to use both Jive and METplus (Brown et al. 2021) as its two primary verification packages. METplus is a verification package that organizes and connects various verification tools developed by the Developmental Testbed Center, with the Model Evaluation Tools (MET) as its core. METplus is increasingly being adopted by meteorological agencies worldwide, facilitating easy and fair comparison of model verification across institutions. Compared to Jive, METplus has superior ability in evaluating NWP forecasts, particularly with its spatial verification capability which Jive does not contain e.g., MODE (Davis et al. 2006), intensity-scale techniques (Casati et al. 2004), HiRA (Mittermaier 2014), and distance mapping methods (e.g., Gilleland et al. 2020).

Conversely, Jive offers several user-focused methods not available in METplus which are suitable for evaluating post-processed and official forecasts disseminated to the public. All metrics are designed to work on multi-dimensional arrays which allows a single implementation of a metric to work on spatial data, data at point locations, or on district data. The verification code in Jive is entirely in Python, whereas METplus is a python wrapper around the MET core, which is written in C++. Most Bureau scientists are familiar with Python with fewer familiar with C++. This means that Bureau scientists can quickly develop and implement new metrics in Jive. Furthermore, metrics in Jive have demonstrated scalability on high-performance computers through parallel computing with Dask (e.g., Loveday et al. 2024).

The technological designs of Jive and METplus reflect the original needs of each package at the time at which each was established. While both systems have naturally developed some overlap as they expanded, it is unrealistic to expect a single verification package to meet all the needs of every meteorological agency. At the Bureau, the strengths of METplus and Jive complement each other, and efforts are being directed towards understanding how they can be optimally used together. At this stage, both tools serve clear and valued purposes. Future work could involve centralizing all forecast and observation data in a single location and ensuring easy accessibility for use by either tool. Additionally, the Bureau is considering enhancing the capability of existing Jive dashboards to display METplus output.

### *Making Jive metrics open source and using Jive as an educational tool*

Since Jive contains several novel methods either unavailable elsewhere or not yet implemented in Python, the Bureau has started migrating the metrics component of Jive to an open-source package called *scores* (Leeuwenburg et al. 2024). *Scores* can serve as a place for the international community to use and add new xarray-based Python implementations of new verification scores. These methods can be imported into other Python verification packages or used as is. *Scores* also contains commonly used metrics so it can be used as a standalone package. Additionally, a Jupyter Notebook tutorial has been created for every verification metric in *scores*, which means that *scores* can be used as an educational tool to learn about verification science.

Metrics in *scores* can serve as a blueprint for implementation in a different language in other verification packages, for example, an implementation in C++ in MET which would support the Research to Operations Pipeline. This prospect demonstrates another way that the Bureau could benefit from focusing on both Jive and METplus.

Jive has supported undergraduate and postgraduate internships. By allowing easy access to a variety of cleaned datasets in a Python environment, it has enabled students to promptly commence the analysis of forecast and observation data (e.g., Short 2020). Tutorial Jupyter Notebooks and recorded training videos have provided Bureau staff with guidance on how to use Jive. Jive has also been integrated into the Bureau's Basic Instruction Package for Meteorologists to aid in instructing graduate meteorologists on forecast verification. Jive data has also been used to support public "hackathons".

## Outlook

Meteorologists will continue to witness a transformation in their roles, especially with the rapid advancements in artificial intelligence (Stuart et al. 2022; Roebber and Smith 2023). Verification will play an ever-increasing role in supporting decisions around automation. This paper introduces the Jive verification system and shows how it helped inform decisions around creating the optimal mix between meteorologists and automated processes. The examples presented demonstrate the important role of verification in meteorological organizations to allow them to tackle questions around automation and the role of the meteorologist. A large part of Jive's success lies in providing all staff with easy access to organized, quality-controlled data, a data analysis environment, and user-focused verification methods. Streamlining the forecast process, while simultaneously improving the quality of forecasts, has only been possible through cross-disciplinary teams involving verification scientists and software developers who worked closely with operational meteorologists to embed verification within the daily forecast process.

Until now, Jive has primarily evaluated public weather forecasts rather than weather warnings. An exception is the heatwave warning verification dashboard. Looking ahead, Jive will play a role in understanding the quality of warnings and informing meteorologists on how to produce better warnings. Evaluating warnings will help the organization clarify how the warning service is defined, support the development of automated warning guidance, and understand how warnings created by meteorologists compare with those created by (still to be developed) future automated sources. One notable challenge will be acquiring or developing suitable observation datasets for warning verification, consistent with the challenges noted by Pagano et al. (2024).

As the Bureau continues to produce more tailored forecasts for diverse forecast users, verification demands will rise, necessitating ongoing development and implementation of new user-focused verification methods. Jive’s utility could expand through the development of dashboards specifically designed to assist meteorologists advising industry, emergency services, and other users to make optimal decisions (e.g., step 6 in Fig 1). To address the growing demand for verification, it will be necessary for operational meteorologists to continue to increase their involvement and use of Jive.

The demonstrated positive impacts of Jive have allowed the Bureau to produce more accurate and valuable weather forecasts for the Australian community. Verification is expected to be a key driver of continuous improvement in forecast services, well into the future.

*Acknowledgments.*

We would like to thank Barbara Casati (ECCC) and two anonymous reviewers for their constructive feedback on this paper. Additionally, we would like to thank Jason West, Benjamin Owen, and Debbie Hudson from the Bureau of Meteorology for feedback on an earlier manuscript. The success of the Jive verification system would not have been possible without the scientists and software developers who contributed to Jive. We would like to thank those who previously worked on Jive at the Bureau of Meteorology, including Michael Foley, Harry Jack, Ioanna Ioannou, Maoyuan Liu, Gary Weymouth, Alexei Hider, Alistair McKelvie, Nandun Thellamurege, Benjamin Price, and Aidan Griffiths. Additionally, managers, project managers, and business analysts have supported the project including Harald Richter, Daria Wickramasinghe, Samhita Barman, and Paul Merrill. We are also grateful to Andrew Hicks and Tracy Rowland from the Application Support team for deploying Jive on operational infrastructure. We would also like to acknowledge operational meteorologists who have worked closely with the project, including Peter Newham, Bradley Wood, Helen Zhou, and Pheobe de Wilt.

*Data Availability Statement.*

The code to generate all the verification results and figures is available at https://github.com/nicholasloveday/jive_paper. Data used to produce figures is available on Zenodo (Loveday 2024). *Scores* code is available at https://github.com/nci/scores with documentation at https://scores.readthedocs.io.

# APPENDIX A

## Summary of forecast and observation data stored in Jive

Forecast data are summarized in Table A1, and observation data is summarized in Table A2.

| **Data source** | **Parameters** | **Description** |
|---|---|---|
| NWP | Surface level forecasts for rainfall, temperature, and cloud coverage. Surface and upper-level forecasts for wind speed on-hour, dewpoint, and wind direction. | Includes ACCESS (Puri et al. 2013) global and city models, ECMWF high-resolution forecast (HRES) |
| GOCF | Surface level forecasts for rainfall, dewpoint, and cloud coverage. Surface and upper-level forecasts for wind speed on-hour, wind direction, and temperature. | Gridded Operational Consensus Forecast (GOCF) (Engel and Ebert 2012) is a bias-corrected "poor man's" ensemble, that is currently the basis for most automated guidance that flows into AutoFcst. |
| IMPROVER | Surface level forecasts for rainfall and cloud coverage. Surface and upper-level forecasts | Integrated Model post-PROcessing and VERification (IMPROVER) (Roberts et |
| | for wind speed on-hour, wind direction, dewpoint, and temperature. | al. 2023) will replace GOCF midway through 2024. |
| Calibrated Thunder | Thunderstorm forecasts. | Probabilistic thunderstorm guidance derived from forecast parameters in ACCESS global ensemble. It is a modified version of the system described in (Bright et al. 2005) |
| AutoFcst | Surface level forecasts for rainfall, wind speed (max-in-hour and on-hour), wind direction, temperature, dewpoint, relative humidity, significant wave height, fire danger indices and fuel load information, and thunderstorms. | AutoFcst is the currently available automated alternative to the official human-curated gridded forecasts (Griffiths and Jayawardena 2022). |
| Official | Surface level forecasts for rainfall, wind speed (max-in-hour and on-hour), wind direction, wind gust, temperature, dewpoint, relative humidity, significant wave height, cloud coverage, fire danger indices and fuel load information, thunderstorms, and heatwave forecasts and warnings. | Gridded forecasts are issued to the Australian Digital Forecast Database (ADFD; BOM 2015) to populate the website, mobile app, and may be used by downstream users. Additionally, some Bureau warning products are stored in the Jive database. |
| Private sector | Surface level forecasts for rainfall, cloud cover, relative humidity, dewpoint, temperature, ultraviolet radiation index, wind speed, wind direction, and wind gust. | Third-party forecasts. |

Table A1. Forecast datasets that are stored in the Jive database. Forecast data includes point-based forecasts at AWS locations and gridded forecasts for various parameters over Australia. Post-processed rainfall forecasts (including the official forecasts) include 3-hourly and daily forecasts of the chance of various rainfall amounts, as well as quantile and expected-value forecasts. Surface temperature forecasts include hourly, daily maximum, and daily minimum forecasts. Upper-level forecasts are for the 950hPa, 900hPa, 850hPa, 800hPa, and 700hPa pressure levels.

| **Data source** | **Parameters** | **Description** |
|---|---|---|
| AWS | Rainfall, temperature, dewpoint, relative humidity, wind speed (on-hour and max-in-hour), wind direction, and fire danger indices. | Data from AWS stations. |
| WZTLN | Lightning count within a 10km radius of the centroid of each grid cell across 3-hour and 24-hour periods. | Lightning data from WeatherZone Total Lightning Network sensors. |
| Wave buoys | Significant wave height. | Data from various buoys, mostly third party. |
| Heatwave observations | Gridded heatwave index observations, and district-based observations. | Australian Water Availability Project (AWAP) grids (Jones et al. 2009) |

Table A2. Observation sources that are stored in the Jive database. While no upper-level observations are included in the Jive database, radiosonde observations can be retrieved in the Jupyter notebooks by connecting to other Bureau databases.

## APPENDIX B

### AWS locations

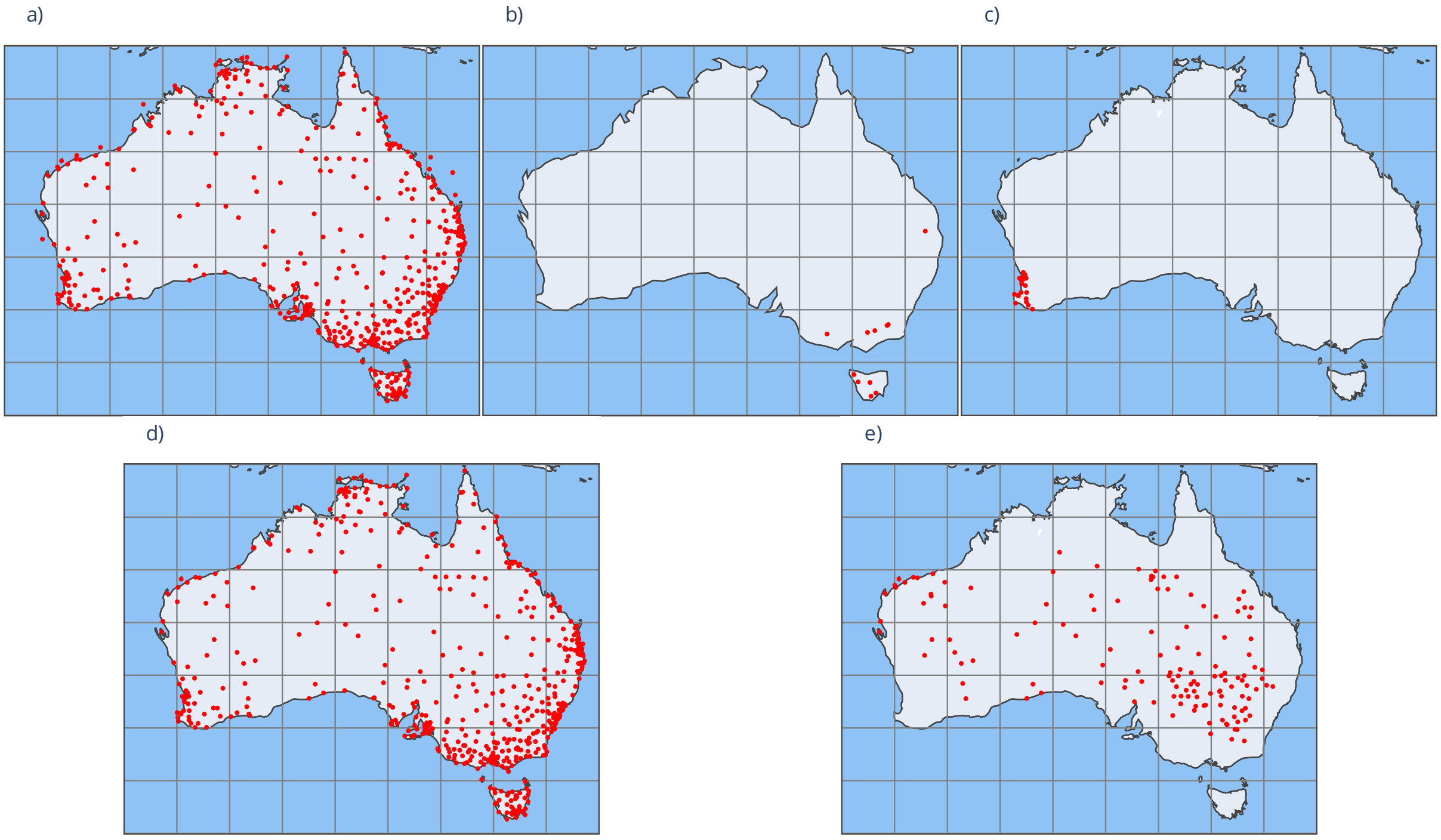

Figure B1. Maps of AWS locations used for verification mentioned in this paper. These station groups are available in Jive and were chosen with operational meteorologists. **a)** Australian wind stations, **b)** High mountain peak stations, **c)**, Southwestern Western Australia stations, **d)** Australian temperature stations over land, and **e)** Inland stations.